# Nuclear dynamics of mass asymmetric systems at balance energy.

Supriya Goyal and Rajeev K. Puri*
*Department of Physics, Panjab University, Chandigarh-160014, INDIA*
* email: rkpuri@pu.ac.in

## Introduction

In the search of nuclear equation of state as well as of nuclear interactions and forces, collective transverse flow has been found to be of immense importance [1]. At low incident energies, the collective transverse flow is dominated by attractive interactions and the flow is expected to be negative, while at high incident energies, the flow is dominated by nucleon-nucleon repulsive interactions and is expected to be positive. While going from low to high incident energies, collective transverse flow vanishes at a particular value of energy, which is termed as *Balance Energy ($E_{bal}$)* [1]. The $E_{bal}$ has been reported to be of significance toward the understanding of nuclear interactions and related dynamics [1,2]. Extensive investigations have been done on symmetric and nearly symmetric reactions in order to calculate the accurate value of Ebal in the last 2-3 decades, both experimentally as well as theoretically [1,2]. Some attempts are also reported that deal with the impact parameter dependence of the $E_{bal}$ [2]. Recently, the role of mass asymmetry of the reaction on the $E_{bal}$ and its mass dependence has been reported [3]. It has been found that for large asymmetries, the effect of mass asymmetry can be 15% with momentum dependent interactions and in the absence it can be 40% [3]. A complete systematic analysis of the nuclear dynamics at the $E_{bal}$ for nearly symmetric reactions is reported in the literature [4], but no attempt has been made to see the effect of mass asymmetry on the nuclear dynamics at balance energy. We aim to address this question in the present study using Quantum Molecular Dynamics (QMD) model [5].

## Model

The QMD model simulates the heavy-ion reactions on event by event basis. This is based on a molecular dynamic picture where nucleons interact via two and three-body interactions. The nucleons propagate according to the classical equations of motion:

$$\frac{d\mathbf{r}_i}{dt} = \frac{dH}{d\mathbf{p}_i} \text{ and } \frac{d\mathbf{p}_i}{dt} = -\frac{dH}{d\mathbf{r}_i}, \quad (1)$$

where H stands for the Hamiltonian which is given by

$$H = \sum_i \frac{\mathbf{p}_i^2}{2m_i} + V^{tot}. \quad (2)$$

Our total interaction potential $V^{tot}$ reads as

$$V^{tot} = V^{Loc} + V^{Yuk} + V^{Coul} + V^{MDI}, \quad (3)$$

where $V^{Loc}$, $V^{Yuk}$, $V^{Coul}$, and $V^{MDI}$ are, respectively, the local (two and three-body) Skyrme, Yukawa, Coulomb and momentum dependent potentials. $E_{bal}$ is calculated by using the quantity "*directed transverse momentum* $<P^{dir}_x>$", which is defined as:

$$\left\langle p_x^{dir} \right\rangle = \frac{1}{A}\sum_{i=1}^{A} sign\{y(i)\} p_x(i). \quad (4)$$

Here $y(i)$ is the rapidity and $p_x(i)$ is the transverse momentum of $i^{th}$ particle. The rapidity is defined as:

$$y(i) = \frac{1}{2}\ln\frac{E(i) + p_z(i)}{E(i) - p_z(i)}, \quad (5)$$

where E(i) and $p_z(i)$ are, respectively, the total energy and longitudinal momentum of $i^{th}$ particle. The $E_{bal}$ was then deduced using a straight line interpolation.

## Results and discussion

The asymmetry of a reaction is defined by the parameter called asymmetry parameter ($\eta$) and is given by:

$$\eta = \left|\frac{A_T - A_P}{A_T + A_P}\right|, \qquad (6)$$

where $A_T$ and $A_P$ are the masses of target and projectile, respectively. The $\eta = 0$ corresponds to symmetric reactions and nonzero values of $\eta$ defines different asymmetries of a reaction. For the present study, we simulated the various reactions at semi-central geometries by varying the asymmetry of a reaction from $\eta = 0.1$-$0.7$, keeping the total mass of the system ($A_{TOT}$) fixed as 240. For the present study, we employed a soft equation of state (K=200 MeV) with momentum dependent interactions along with energy dependent cugnon cross-section. In fig. 1, we display the maximal value of the average density, maximum density, total number of the allowed collisions (obtained at the final stage), and the final saturated spectator and participant matter as a function of $\eta$. Lines are just to guide the eye. From the figure, it is clear that mass asymmetry of reaction has significant effect on the nuclear dynamics. With increase in $\eta$, all quantities showed in the Fig. 1 decreases except spectator matter. One should note that with increase in $\eta$ the balance energy increases [3]. This is due to the decrease in overlap region which leads to decrease in nucleon-nucleon collisions (shown clearly in Fig. 1(c)). The maximum density is found to be less effected by change in the mass asymmetry. The spectator matter increases while the participant matter decreases with increase in $\eta$.

## Acknowledgments

This work is supported by a research grant from the Council of Scientific and Industrial Research (CSIR), Govt. of India, vide grant No. 09/135(0563)/2009-EMR-1.

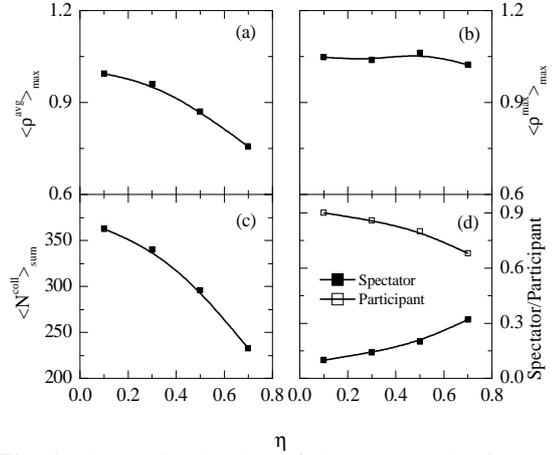

**Fig. 1** The maximal value of the average density (a), maximum density (b), total number of the allowed collisions (obtained at the final stage) (c), and the final saturated spectator and participant matter (d) as a function of $\eta$.